\documentclass[12pt,preprint]{adap}
\usepackage{amssymb}
\usepackage{fancyhdr}
\usepackage{mathptmx} 

\usepackage{natbib}
\usepackage{graphics,graphicx}

\def\apjl{\rm ApJL}
\def\apjs{\rm ApJS}

\def\aap{\rm A\&A}

\def\checkbox{\square\!\!\!\!\!\!\!{\bf X}}

\usepackage{color}
\usepackage{wrapfig}

\usepackage[compact]{titlesec}
\titlespacing{\section}{0pt}{6pt}{4pt}
\titlespacing{\subsection}{0pt}{6pt}{4pt}
\titlespacing{\subsubsection}{0pt}{6pt}{4pt}

\newcommand{\msun}{M$_{\odot}$}

\begin{document}


%
%





\thispagestyle{empty}
\begin{center}

{\large\bf Astro2020 Science White Paper}

\bigskip{\Large\bf Where are the Intermediate Mass Black Holes?}

\end{center}

\vfill

\noindent{\bf Thematic Areas:}
$$\vbox{\hsize 5truein \halign{# \hfil & #\hfil \cr
$\square$ Planetary Systems & $\square$ Star and Planet Formation \cr
$\checkbox$ Formation and Evolution of Compact Objects & $\square$ Cosmology and Fundamental Physics \cr
$\square$  Stars and Stellar Evolution & $\square$ Resolved Stellar Populations and their Environments \cr
$\checkbox$ Galaxy Evolution & $\checkbox$ Multi-Messenger Astronomy and Astrophysics \cr
}}$$

\vfill

\begin{center}
{\it by}

Jillian Bellovary
\\
City University of New York, Queensborough Community College, 222-05, 56th Avenue, Bayside, NY 11364, USA \\
and 
American Museum of Natural History, Central Park West at 79th Street New York, NY 10024-5192, USA
\\
{\tt jbellovary@amnh.org}

\bigskip{\it in collaboration with}

Alyson Brooks$^1$, Monica Colpi$^2$, Michael Eracleous$^3$, Kelly Holley-Bockelmann$^{4,5}$, Ann Hornschemeier$^6$,  Lucio Mayer$^7$, Priya Natarajan$^8$,  Jacob Slutsky$^6$, \& Michael Tremmel$^8$ 

\bigskip{\it and with thanks to}

Jenny Greene$^9$, M. Coleman Miller$^{10}$, and Amy Reines$^{11}$ 

for a critical reading of the paper and very helpful comments and suggestions.

\end{center}


\vfill\vbox{\noindent\small
$^1$Department of Physics \& Astronomy, Rutgers, the State University of New Jersey, USA\\
$^2$Department of Physics, University of Milano Bicocca, Italy \\
$^3$Department of Astronomy \& Astrophysics, The Pennsylvania State University, USA \\ 
$^4$Department of Physics \& Astronomy, Vanderbilt University, USA\\
$^5$Department of Physics, Fisk University, USA\\
$^6$NASA, Goddard Space Flight Center, USA \\
$^7$Center for Theoretical Astrophysics and Cosmology, University of Zurich, Switzerland \\
$^8$Department of Astronomy, Yale University, USA \\
$^9$Department of Astrophysical Sciences, Princeton University, USA\\
$^{10}$Department of Astronomy and Joint Space-Science Institute, The University of Maryland, USA \\
$^{11}$Department of Physics, Montana State University, USA \\
}
%


\clearpage
\setcounter{page}{1}


\setlength{\parindent}{0.25in}

{\noindent
{\bf Executive Summary:}
Observational evidence has been mounting for the existence of intermediate mass black holes (IMBHs, $10^2$--$10^5\;$\msun), but observing them at all, much less constraining their masses, is very challenging.  In one theorized formation channel, IMBHs are the seeds for supermassive black holes in the early universe.  As a result, IMBHs are predicted to exist in the local universe in dwarf galaxies, as well as wandering in more massive galaxy halos.  However, these environments are not conducive to the accretion events or dynamical signatures that allow us to detect IMBHs.  The Laser Interferometer Space Antenna (LISA) will demystify IMBHs by detecting the mergers of these objects out to extremely high redshifts, while measuring their masses with extremely high precision.  These observations of merging IMBHs will allow us to constrain the formation mechanism and subsequent evolution of massive black holes, from the `dark ages' to the present day, and reveal the role that IMBHs play in hierarchical galaxy evolution.
}







\section{Introduction: Are there intermediate mass black holes?}
Once thought of as a hypothetical oddity, the case for the existence of intermediate mass black holes (IMBHs) has been building in recent years, both observationally and theoretically.  The previously assumed `mass gap' between stellar black holes ($\sim$1--100~\msun) and supermassive black holes (SMBHs; $\sim$$10^6$--$10^9\;$\msun) may be filled in, though to what extent is still very uncertain.  Motivated by theoretical predictions for the high-redshift formation of IMBHs as SMBH `seeds,' various observations have identified several candidates in the IMBH mass range ($10^2$--$10^5$~\msun).

The idea of IMBHs as SMBH seeds is appealing because of the sheer difficulty in creating a $10^9$~\msun\ SMBH in less than one billion years, as seen with observations of high redshift quasars \citep[e.g.][]{Banados17}.  A stellar-mass black hole can only achieve such large mass if it accretes at the Eddington limit for its entire life up to that point \citep{Haiman01}.  In contrast, a more massive IMBH seed gives black holes a head start, allowing them to grow to supermassive sizes much more quickly \citep{Haiman01, Volonteri03}.   The primary theoretical models for creating such objects invoke the remnants of Population III stars, direct collapse of massive pristine gas clouds, and the runaway collision of objects in dense star clusters. Regardless of which (if any) of these mechanisms is responsible for seed SMBHs, any/all of them can create IMBHs, and there may be multiple pathways to IMBH formation.



\vspace{-1mm}
\subsection{Theoretical Motivations: Seeds of Supermassive Black Holes}

Several theories have been postulated for the origin of SMBH seeds.  Some of the leading candidates are:
(a) Population III star remnants (100--1000 \msun, form at $15 < z < 30$), (b) Direct collapse black holes ($10^4$--$10^6\;$\msun, form at $10 < z < 20$), and (c) 
Collapsing stellar clusters ($\sim$$10^3\;$\msun, form at high or low $z$).
Each of these processes results in a different characteristic mass and redshift of formation.  In addition, the efficiencies of seed formation are different with each model.  For example, Population III stars are expected to be common in the early universe, forming in every galaxy in its first epochs of star formation. 
Direct collapse black holes, in contrast, are predicted to be more rare, because they require a special set of conditions in order to form. 
The efficiency of the formation of the primary mechanism for seeding SMBHs will have repercussions at both high and low redshifts, regarding both the occupation fraction of SMBHs in galaxies (particularly low-mass galaxies) as well as the SMBH merger rate.  The different predictions of each model regarding mass, redshift, and efficiency of formation are difficult to differentiate with traditional electromagnetic (EM) observations, but the advent of low-frequency gravitational wave observatories (e.g., LISA, the Laser Interferometer Space Antenna) will usher in a new era of high-precision measurements of black hole masses encoded in the waveforms of gravitational waves (GWs).

\subsection{Observational Motivations: The Evidence is Growing}
The theoretical advances mentioned above have motivated the search for IMBHs in various environments.  If seed formation is truly a high-$z$ event that occurs in fairly small halos, some of these small halos will remain small throughout their lifetimes and exist in the local universe as dwarf galaxies.  Recent works have unveiled evidence for SMBHs in dwarf galaxies \citep{Greene07,Barth08,Reines13,Moran14,Baldassare18}, some of which cross into the IMBH mass range.  The galaxy RGG 118 hosts the smallest known SMBH with a mass of $5 \times 10^4$ \msun, putting it squarely in the IMBH mass range \citep{Baldassare15}.  

Another avenue for hunting down IMBHs exists in off-center compact sources in large galaxies.  The process of hierarchical merging results in the tidal disruption of low-mass galaxies as they enter larger halos.  If those low-mass galaxies host IMBHs, the black holes will wander in the larger galaxy as they inspiral toward the center \citep{Holley-Bockelmann10,Bellovary10,Tremmel17}.  Our own Milky Way hosts such postulated objects; gas clouds near the Galactic Center possessing unusually large velocity dispersions are thought to harbor hidden IMBHs \citep[e.g.][]{Takekawa19}.  If such objects were to experience a rapid accretion event, they would be seen as (likely off-nuclear) ultra-luminous X-ray sources (ULXs).  ULXs are common in star-forming regions and are often attributed to neutron stars or stellar black holes experiencing Eddington or super-Eddington accretion, or beamed emission. 
However, there are a handful of objects which appear to be simply too luminous for such an explanation to be likely.   
The most promising candidate for this scenario is the object HLX-1, an off-nuclear source with a luminosity of $10^{42}$ erg s$^{-1}$, with episodic bursts \citep{Farrell09}. 
This object is likely the nucleus of a small galaxy merging with the larger disk galaxy ESO 243--49, and perturbations from the merger triggered a recurring accretion event onto the IMBH.


IMBH and host galaxy co-evolution can also be studied using observed scaling relations, such as the $M_{\rm BH}$--$\sigma$,   $M_{\rm BH}$--$M_*$,  and $M_{\rm BH}$--$M_{\rm bulge}$ relations and others.  While SMBHs and their host galaxy properties exhibit a tight correlation at higher masses,  the low-mass end is sparsely populated and may exhibit large scatter \citep{Reines15,vandenbosch16}.  As more galaxies hosting IMBHs are discovered, the relation may be filled in, and we will better understand how massive black holes and their hosts grow and evolve.  The shape of the relation can also hint at the formation mechanism for seeds, which could result in a high-mass ``plume'' for direct collapse seeds, or a low-mass concentration for Population III seeds \citep{VolonteriNatarajan}.

While there is disputed dynamical evidence for compact massive objects in the centers of globular clusters which could be IMBHs \citep[e.g.][]{Noyola10,Lutzgendorf13}, the lack of corresponding X-ray or radio signatures hints that even if an IMBH is present, there is no gas to accrete, and thus it is not possible to strongly confirm their existence \citep{Strader12,Wrobel15}.   For a more thorough discussion on the observational evidence for IMBHs, see the excellent review by \citet{Mezcua17}.

\vspace{-1mm}
\section{Challenges for Observations}
\vspace{-1mm}
One major challenge for observations of IMBHs is dynamical.  The radius of influence of a black hole scales linearly with its mass; thus it is difficult to detect the influence of IMBHs on nearby stars without extremely high spatial resolution.  While attempts at measuring IMBH masses in nuclear star clusters have been made \citep{Neumayer12,Nguyen18}, a census can be taken only for the nearest objects.  Upcoming 30-meter class telescopes will greatly expand the number of dynamical mass measurements for nearby black holes, including IMBHs.

Additionally, IMBHs exist mainly in environments where they will not be undergoing strong accretion.  Dwarf galaxies are often irregular in shape, lacking clear centers.  They also tend to exhibit cored density profiles, which can result in an IMBH existing anywhere in the general vicinity of the central region, rather in the exact center.  Gas in dwarf galaxies is clumpy and in many regions has low surface density.  Much of the gas co-located with the IMBH may not be available for efficient accretion.  Recent work has shown that the fraction of dwarf galaxies hosting active galactic nuclei (AGN) is less than a few percent \citep{Reines13,Moran14,Sartori15,Pardo16}.  However the occupation fraction of IMBHs in these galaxies can be much higher.  Low-mass ($\sim$$10^{10}\;$M$_*$) early-type galaxies in the AMUSE-VIRGO survey have a SMBH occupation fraction of  $>20\%$ \citep{Miller15}.   
IMBHs wandering in more massive galaxy halos are even more difficult to detect directly.  These objects could be observable if they still carry remnants of their host, i.e. part of the galaxy (as in HLX-1), a nuclear star cluster, or enough gas to accrete as it traverses the halo. 

At high redshift, the epoch of seed formation is very challenging to observe.  Predicted to occur at $10 < z < 30$, only the lowest-redshift end of this range is thought to be observable in light.  The expected signatures of direct collapse black hole formation include strong Lyman $\alpha$ emission, red colors due to strong obscuration, and X-rays due to gas accretion.  One very luminous Lyman $\alpha$ emitter, known as CR7 at a redshift of $z \sim 6.6$, was thought to be a direct collapse black hole candidate \citep{Agarwal16}, but 
follow-up observations found it is bluer with higher metallicity than previously thought \citep{Bowler17}.  The upcoming James Webb Space Telescope (JWST) will be able to see more distant and redder objects, and may reveal other direct collapse candidates \citep{Natarajan17}, but confirmation of their true nature will be tricky.  

Gravitational waves are a promising avenue for IMBH detection, since they are independent of a black hole's accretion status, but rather depend on its binarity.  GWs from merging black holes have been detected by LIGO \citep[e.g.][]{LIGO}, but the frequency range for merging IMBHs is lower than that which LIGO can detect, unless they are at $z<0.1$.  However, the upcoming LISA mission is optimized for detecting merging black holes in precisely the $10^4$--$10^7\;$\msun\ range.  The LISA Mission, led by the European Space Agency with international partnerships including NASA, will survey the mHz gravitational wave band using a triangular constellation of spacecraft.  The basic principle of the measurement is the same as for LIGO, although unique technical challenges are present for space missions. The recently completed LISA Pathfinder mission and the recently launched Grace Follow-On mission have already proven that the key precision LISA measurement technologies meet or exceed their required performance.


\vspace{-1mm}
\section{Discovering IMBHs in The Early Universe}
\vspace{-1mm}
Observing the epoch of formation of IMBHs at high redshift will become possible with facilities that will soon become available. However, advancements in EM observing technology can only take us so far.  Even if we do observe the birthplaces of IMBHs, confirming what we are seeing will not be trivial.  Separating black hole activity from star formation is challenging whenever a host is unresolved, as these hosts will be.  For example, the red colors expected from direct collapse formation may be indistinguishable from heavily obscured star formation.  Co-spatial X-ray emission may provide supporting evidence for black hole activity, but the mass of the black hole cannot be determined with much accuracy.  As EM observatories improve their sensitivity, identifying IMBH candidates will become easier, however confirming whether an object is an IMBH is going to remain extremely challenging.

\begin{figure}[htb!]
\includegraphics[width=0.45\textwidth]{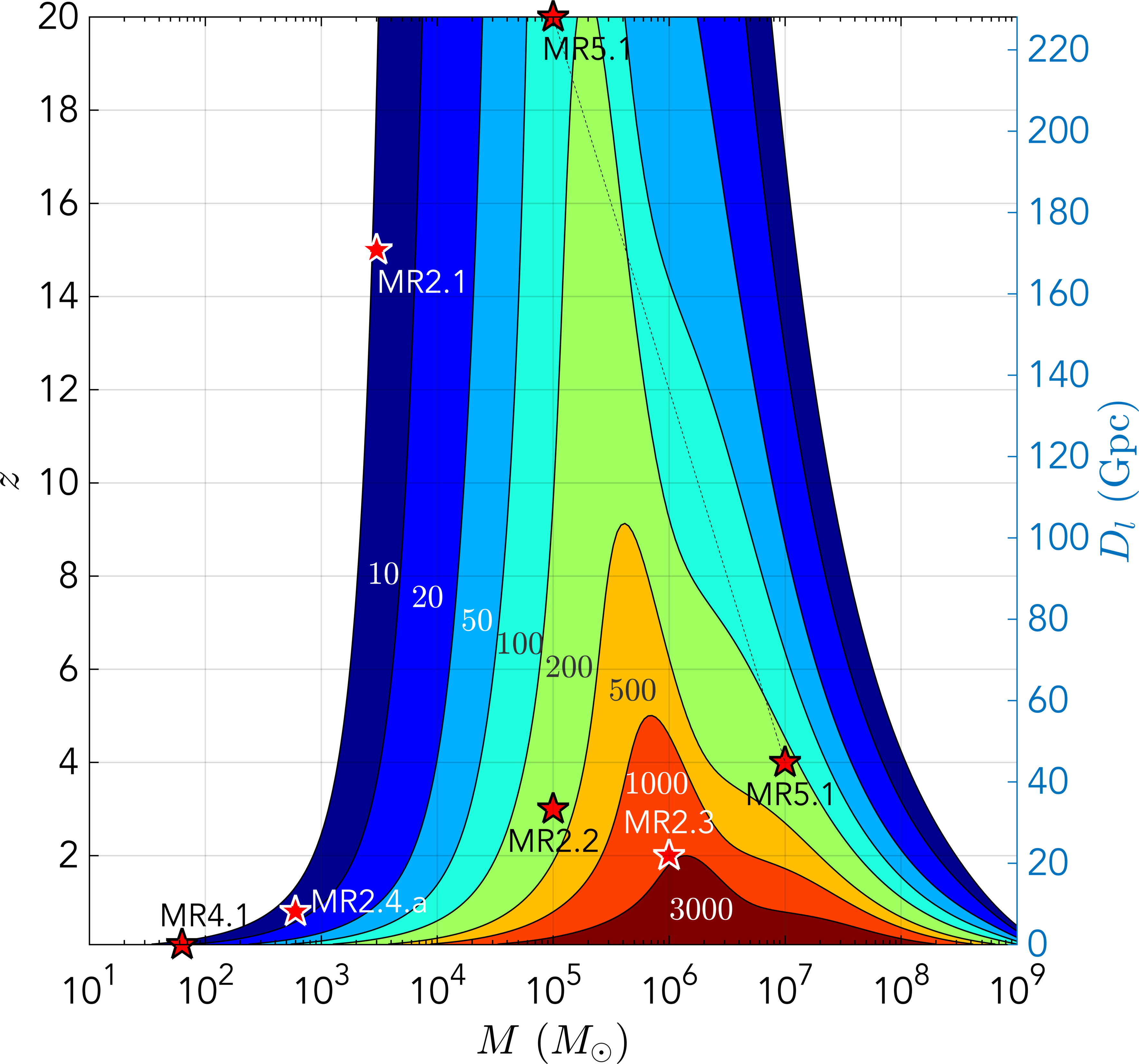}
\includegraphics[width=0.5\textwidth]{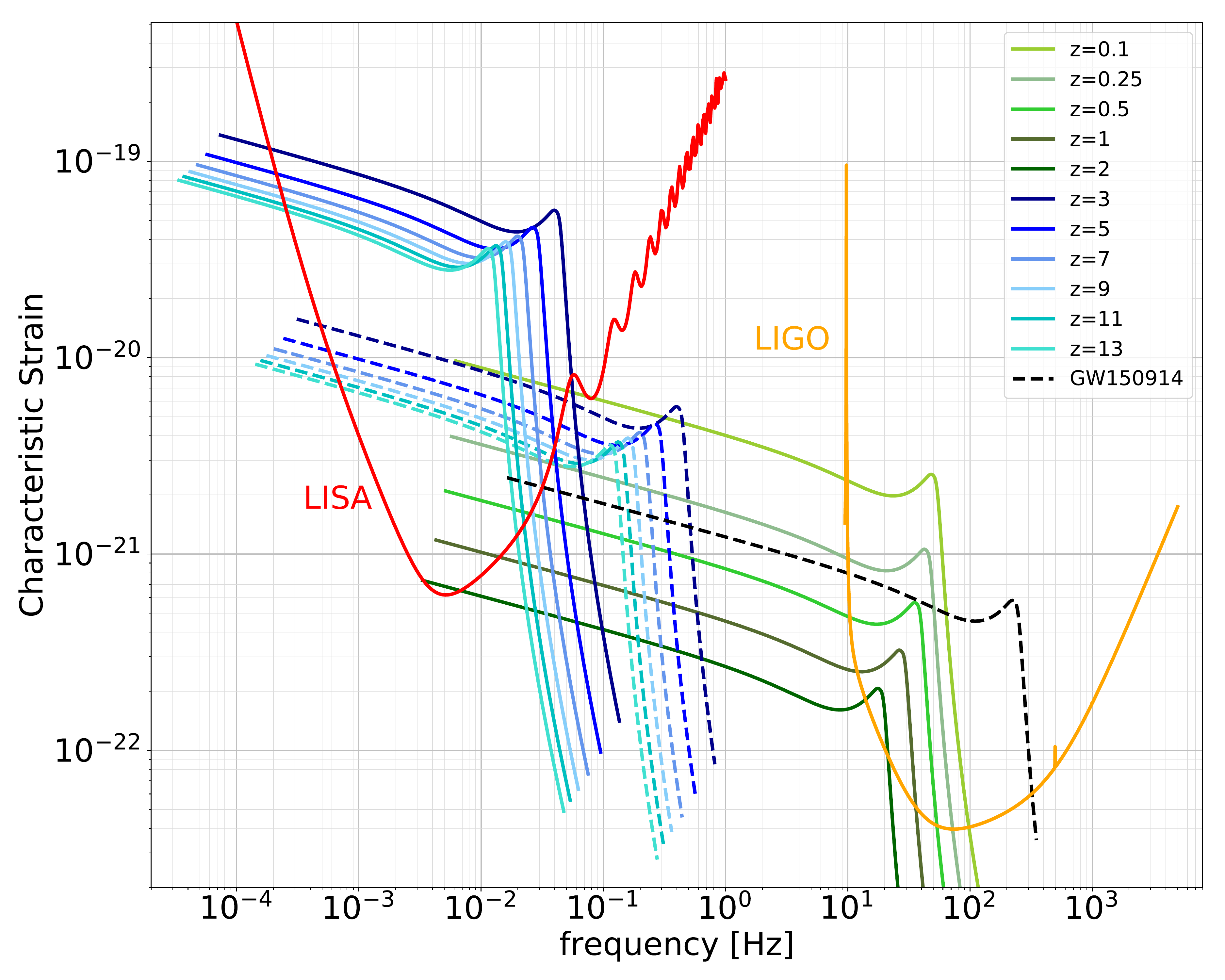}
\vspace{-3mm}
\caption{\small {\em Left panel:}  LISA detection abilities, plotted as redshift on the left axis (and luminosity distance on the right axis)) vs total mass of a SMBH binary.  Colors represent the signal-to-noise ratio.  Labelled stars depict example mass ratios at various redshifts.  Mergers involving IMBHs will be detectable with modest mass ratios all the way out to $z\sim20$ and beyond.  {Figure credit: ESA \citep{LISA}}   {\em Right panel:}  Characteristic strain vs frequency of a variety of IMBH mergers.  The red curve is the LISA sensitivity, while the yellow is LIGO.  Each set of curves represents the inspiral and merger of IMBHs at a range of masses and redshifts (solid blue $10^5$\msun, dashed blue $10^4$\msun, solid green $10^3$\msun).  (Figure courtesy of Colpi-Mangiagli)  \label{fig:waterfall}
}
\end{figure}

LISA can determine masses directly with high accuracy.  
LISA can detect high-$z$ black hole mergers with masses of $10^4$--$10^5\;$\msun\ with signal-to-noise ratios of up to 100 (Figure \ref{fig:waterfall}, left panel).  Masses, mass ratios, and redshift are inherent in the detected waveforms, and can be determined with a precision of 5\%, depending on the effect of lensing.  
LISA will be able to detect coalescing black holes in the (total binary) mass range of $10^3$--$10^7$ \msun\ up to and beyond a redshift of $z = 20$ \citep{LISA}.   To be more precise, the right panel of Figure \ref{fig:waterfall} shows how binaries with total mass $10^5$\msun\ will inspiral and merge within the LISA sensitivity range.  Less massive binaries ($10^4$\msun) will inspiral in the LISA bandwidth, allowing for the characterization of merger properties, but the merger itself will occur at a  frequency above LISA's best sensitivity.  Lower-mass mergers still ($10^3$ \msun) will begin their inspiral in the LISA sensitivity range, and coalesce in the LIGO range, allowing for exciting multi-band detections and characterizations of sources.  


\vspace{-3mm}

   
The rate of mergers will be critical to telling us about the efficiency of IMBH formation.   Since the various seed models are expected to produce IMBHs in different numbers, in different places, and at different times, their merger history can reveal their formation mechanisms.  
Mergers of IMBHs created through different processes are probably likely, and must also be considered.  The efficiency of each mechanism of black hole formation will be thoroughly tested with the rates, redshifts and masses of mergers seen by LISA.

The sites of IMBH mergers are potentially observable with the state-of-the-art observatories that will exist in the 2030s.  Space observatories such as the proposed LUVOIR mission are planned to image more distant and smaller sources than ever before, and are eager and well-suited to image the EM counterparts to GW events.  Other observatories in the planning stages may be able to peer into the distant universe and observe these events as well.  Possible electromagnetic observational synergies include Lynx (X-rays), the Origins Space Telescope (OST, mid-infrared), and the Habitable Exoplanet Imaging Mission (HabEx, optical/near infrared).  
 
LISA's work in determining the dominant formation mechanism of SMBH seeds will help these observatories target such regions specifically, aiming to observe these processes possibly for the first time.    Additionally, because IMBH binaries enter the LISA frequency range months to years before they actually merge, LISA can determine the masses and redshifts of merging objects with enough lead time to search for EM counterparts in targeted regions of the sky.  As a result, we may be able to view IMBH-IMBH mergers as they happen with both EM and GW radiation.

 \vspace{-1mm}
\section{Discovering IMBHs in The Local Universe}
\vspace{-1mm}
While seed formation is expected to be a predominantly high redshift phenomenon, we expect other types of events to be seen by LISA .  The existence of wandering IMBHs in high-mass galaxies will create a reservoir of IMBH-SMBH mergers waiting to happen.  While some IMBHs will have dynamical friction timescales longer than a Hubble time, others will have orbits that decay and bring them into the centers of their hosts.  These mergers can happen at any redshift, including locally.  If the IMBHs have a mass range of 100--$10^5$ \msun, and SMBHs have a mass range of $10^6$--$10^{10}\;$\msun, we expect the full range of mass ratios for these types of events to be 0.1 up to $10^{-8}$.\ 
  Such events are known as extreme or intermediate mass ratio inspirals (EMRIs or IMRIs) and may be the most common form of massive black hole merger \citep{Holley-Bockelmann10,Bellovary10}.   EMRI/IMRI events may also be caused by the mergers of IMBHs which form within AGN accretion disks.  Stellar black holes may rapidly gain mass via migration traps \citep{Bellovary16}, forming IMBHs of 100-1000\msun\ which may merge with their neighboring SMBH \citep{McKernan14}.  These events probably do not have an EM signature, but they have a very characteristic GW waveform which LISA will easily detect.

 
Another predicted but as yet unobserved phenomenon which would unveil IMBHs is a tidal disruption event (TDE).  The tidal disruption of a star by a SMBH gives an easily identifiable signature, and allows one to estimate the mass of the SMBH \citep[for further details on TDEs see review by][]{Komossa15}.  TDEs can also be caused by IMBHs, which in large numbers can provide an independent census of such objects. 
A specifically interesting case is the tidal disruption of a white dwarf by an IMBH.  The combination of the high density of a white dwarf and the smaller event horizon of an IMBH means that an IMBH with mass $\sim 10^5$\msun~ is the largest black hole that could tidally disrupt a white dwarf outside of the event horizon \citep{Sesana08}.  The spectroscopic trademark of this event would be the existence of carbon and oxygen emission lines with a lack of hydrogen and helium lines \citep{Clausen11}.  In addition to the optical detection, an IMBH-white dwarf TDE would be detected by LISA as an EMRI.  A multi-messenger confirmation with two independent IMBH mass measurements will be a phenomenal confirmation of the presence of IMBHs in galaxies.
 
 
 \vspace{-1mm}
   \section{The Bottom Line: IMBHs are Elusive but Important}
\vspace{-1mm}
   
Intermediate mass black holes are not only the missing link between stellar mass and supermassive black holes, but are a key prediction of seed SMBH formation that should exist as relics to this day.  Fully understanding the relationship between SMBHs and their hosts depends on deciphering the masses, number densities, and merger rates of IMBHs.  LISA will be paramount in granting us this insight, as it can detect IMBH mergers at redshifts up to $z \sim 20$, which are known as the `dark ages' of the universe, opaque to EM radiation.  

LISA will provide mass and redshift distributions of IMBH mergers with unprecedented accuracy,  allowing not only the determination of the frequency of mergers, but also the initial masses of seed black holes. In addition, this information will potentially allow us to untangle whether gas accretion or mergers with other IMBHs are the predominant form of growth for SMBH seeds.  The growth mechanism may evolve with black hole mass as well as redshift, and knowing how often IMBHs merge may be the only insight we can obtain from the earliest growth epoch.


Overall, IMBHs are a fundamental key to understanding the formation, growth, and evolution of SMBHs and of the overall black hole population in the universe.  LISA's ability to detect IMBH mergers across most of cosmic time, combined with the prowess of EM observations, will be sure to answer crucial questions as well as pose new ones.

\newpage
\bibliography{imbh}

\end{document}